\documentclass[secnumarabic, graphics,floatfix,nofootinbib,
nobibnotes,aps,prb,superscriptaddress,preprint]{revtex4-1}

\usepackage{amssymb}
\usepackage{amsmath}

\usepackage{graphicx}
\usepackage{natbib}

\begin{document}

\title{ Magnetoelectric Effects in the Spiral Magnets CuCl$_{2}$ and CuBr$_{2}$}

\author{P. Tol\'{e}dano}
\thanks{Corresponding author}
\email{pierre.toledano@wanadoo.fr}

\affiliation{Laboratoire de Physique des Syst\`{e}mes Complexes, Universit\'{e} de Picardie, 80000 Amiens, France}
\author{A. P. Ayala}
\affiliation{Departamento de F\'{\i}sica, Universidade Federal do Cear\'{a}, 60440-970, Fortaleza, Brasil}
\author{A. F. G. Furtado Filho}
\author{J. P. C. do Nascimento}
\author{M. A. S. Silva}
\author{A. S. B. Sombra}
\affiliation{Departamento de F\'{\i}sica, Universidade Federal do Cear\'{a}, 60440-970, Fortaleza, Brasil}
\affiliation{LOCEM, Universidade Federal do Cear\'{a}, 6040-970, Fortaleza, Brasil}

\date{\today}

\begin{abstract}
The nature and symmetry of the transition mechanisms in the spin-spiral copper halides CuCl$_2$ and CuBr$_2$ are analyzed theoretically. The magnetoelectric effects observed in the two multiferroic compounds are described and their phase diagram at zero and applied magnetic fields are worked out. The emergence of the electric polarization at zero field below the paramagnetic phase is shown to result from the coupling of two distinct spin-density waves and to be only partly related to the Dzialoshinskii-Moriya interactions. Applying a magnetic field along the two-fold monoclinic axis of CuCl$_2$ yields a decoupling of the spin-density waves modifying the symmetry of the phase and the spin-spiral orientation. The remarkable periodic dependences of the magnetic susceptibility and polarization, on rotating the field in the monoclinic plane, are described theoretically.
\end{abstract}


\maketitle

\section{Introduction}

Multiferroic properties in helimagnets are found in an increasing variety of magnetic materials \cite{Fiebig2005,Eerenstein2006a,Kimura2007a}. Although the emergence of ferroelectricity was initially found in incommensurate antiferromagnetic phases of oxides, this property was then evidenced in commensurate antiferromagnets \cite{Jodlauk2007,Toledano2011} and more recently in the non-oxide compounds CuCl$_{2}$ and CuBr$_{2}$ \cite{Seki2010,Banks2009,Lee2012,Zhao2012}. However, in contrast with multiferroic phases in oxides which generally arise in a two-step sequence of phases, below the region of stability of a non-polar helical phase, in CuCl$_{2}$ and CuBr$_{2}$ ferroelectric properties coexisting with a spin-spiral antiferromagnetic order are observed directly below the paramagnetic phase. Another specific feature of multiferroicity in these compounds is the sensitivity of their magnetic structure under applied field and the complexity of their magnetoelectric properties. Here, we analyze theoretically the nature and symmetry of the transition mechanism in the two copper halides, their phase diagram at zero and applied magnetic field, and their observed magnetoelectric effects.

\section{Phase diagram at zero field}

\begin{figure}[ht]
  \centering
  \includegraphics[width=5in]{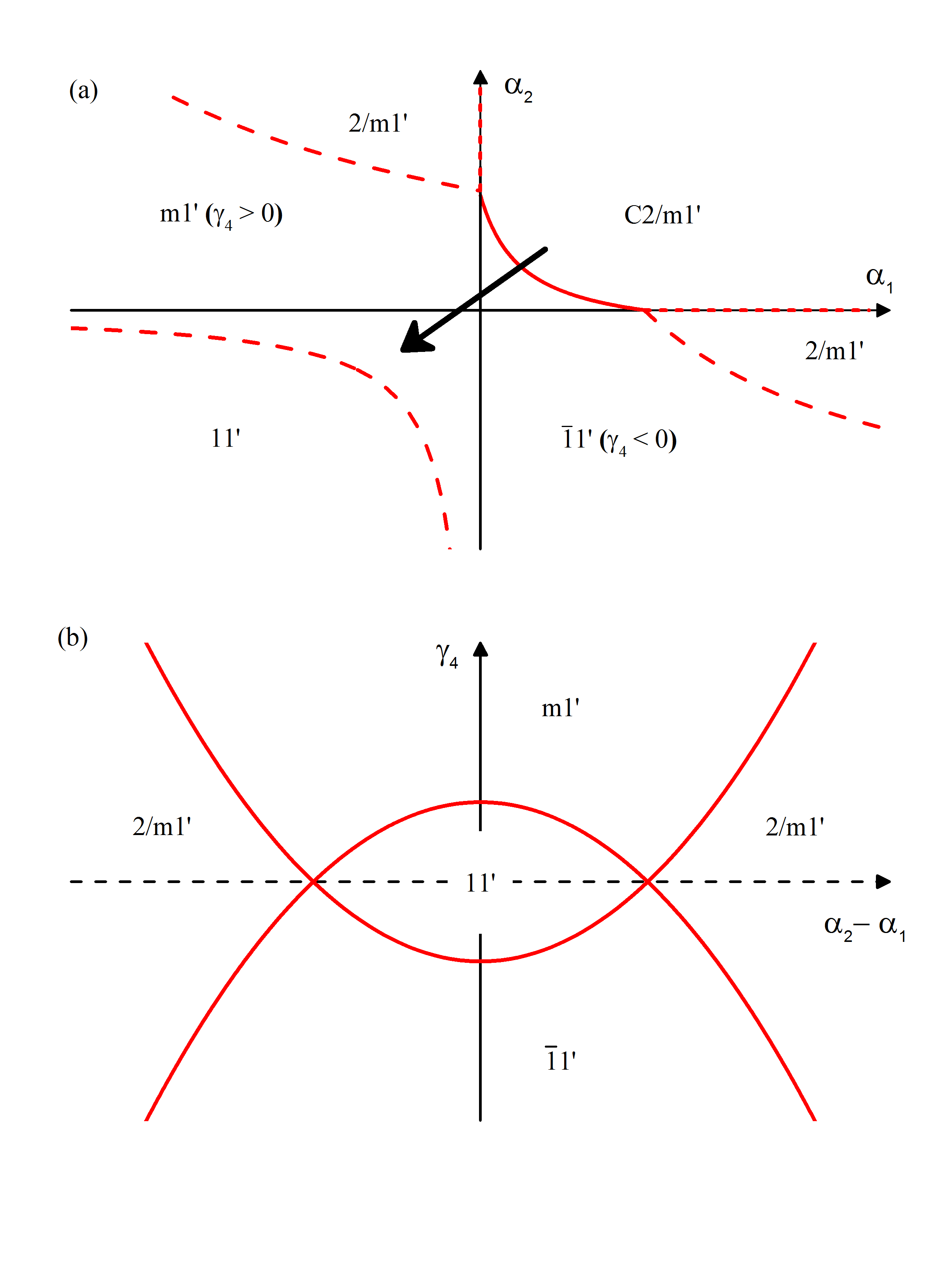}
  \caption{Theoretical phase diagrams at zero fields in the (a) ($\alpha _{1} ,\alpha _{2} $)  and (b) $(\gamma _{4} ,\alpha _{2} -\alpha _{1} )$ planes assuming a eighth-degree expansion for the free-energy given by Eq. \eqref{EQ1}. All solid red curves in (b) and hatched red curves in (a) are second-order transition lines, whereas the solid red curve in (a) denotes a first-order transition line. Fig. (a) shows that the paramagnetic $C2/m1'$ to polar $m1'$ phase transition should occur across the region of stability of a non-polar $2/m1'$ phase or directly across a first-order transition as observed experimentally.
  }\label{fig1}
\end{figure}

The incommensurate wave-vector \textbf{k = }(1, k$_{y}$, 0.5) arising at the transition to the spin-spiral phase, which takes place at T$_{N }$= 24 K in CuCl$_{2}$ (k$_{y}$ =0.226) \cite{Banks2009} and T$_{N}$ = 73.5 K  in CuBr$_{2}$ (k$_{y}$ = 0.235)\cite{Zhao2012} is located at the surface (R-line) of the monoclinic C Brillouin-zone. It is associated with two bi-dimensional irreducible representations \cite{Kovalev1965} $\Gamma _{1} $ and $\Gamma _{2} $ of the paramagnetic space-group $C2/m1'$, the matrices of which are given in Table \ref{table1}. Denoting $\eta _{1} =\rho _{1} \cos \theta _{1} ,\eta _{2} =\rho _{1} \sin \theta _{1} $ and $\zeta _{1} =\rho _{2} \cos \theta _{2} ,\zeta _{2} =\rho _{2} \sin \theta _{2} $ the order-parameter components respectively associated with $\Gamma _{1} $ and $\Gamma _{2} $, the Landau free-energy corresponding to the reducible representation $\Gamma _{1} +\Gamma _{2} $ reads:

\begin{equation} \label{EQ1}
F=\frac{\alpha _{1} }{2} \rho _{1}^{2} +\frac{\beta _{1} }{4} \rho _{1}^{4} +\frac{\gamma _{1} }{6} \rho _{1}^{6} +\frac{\alpha _{2} }{2} \rho _{2}^{2} +\frac{\beta _{2} }{4} \rho _{2}^{4} +\frac{\gamma _{2} }{6} \rho _{2}^{6} +\frac{\gamma _{3} }{2} \rho _{1}^{2} \rho _{2}^{2} +\frac{\gamma _{4} }{2} \rho _{1}^{2} \rho _{2}^{2} \cos 2(\theta _{1} -\theta _{2} ) + \frac{\gamma _{5} }{4} \rho _{1}^{4} \rho _{2}^{4} \cos ^{2} 2(\theta _{1} -\theta _{2} )
\end{equation}

where $\alpha _{1} =a_{1} (T-T_{c1} ),\alpha _{2} =a_{2} (T-T_{c2} )$ and  $a_{1} ,a_{2} ,\beta _{1} ,\beta _{2} ,\gamma _{i} \; (i=1-5)$ are phenomenological constants. Table \ref{table2} lists the symmetries of the stable phases that may arise below the paramagnetic phase at zero field deduced from a Landau symmetry analysis which requires considering an eighth degree term in $F$ for stabilizing the full set of low symmetry phases \cite{Toledano1987}. Because of the incommensurate character of the phases the symmetries correspond at zero field to \textit{grey magnetic point groups} \cite{Dvorak1983}. Column (d) of Table II indicates the spontaneous components of the electric polarization emerging in each phase. One can verify that the P$^{a}$ and P$^{z}$ polarization components, observed experimentally at zero field in CuCl$_{2}$, is induced by the reducible representation $\Gamma _{1} +\Gamma _{2} $ for the equilibrium values of the order-parameter components $\rho _{1} \ne 0,\rho _{2} \ne 0,\theta _{1} -\theta _{2} =(2n+1)\frac{\pi }{2} $ , or equivalently  $\eta _{1} \ne 0,\eta _{2} =0,\zeta _{1} =0,\zeta _{2} \ne 0$. It coincides with a phase of symmetry $m1'$ which allows a spin-spiral in the b-c plane. Figs. \ref{fig1}(a) and \ref{fig1}(b) show the theoretical phase diagrams involving the phases listed in Table II and the thermodynamic path corresponding to the paramagnetic to multiferroic phase.

\begin{table}[h]
  \caption{Irreducible representations $\Gamma _{1} $and $\Gamma _{2} $ of the paramagnetic space-group $C2/m1'$ associated with the wave-vector \textbf{k}=(1, k$_{y}$, 0.5).
 T is the time reversal operator. The matrices of $\Gamma_1$ and $\Gamma_2$ are given in complex form }\label{table1}
  \centering
  \begin{tabular}{cccccccc}
  \\
    $C2/m1'$ & $\left(2_{Y}\left|000\right.\right)$ & $\left(\bar{1}\left|000 \right.\right)$ & $\left(m_{ac} \left|000\right. \right)$ & $T$ & $\left(1\left|0b0\right. \right)$  & $\left(1\left|\frac{a}{2} 0\frac{c}{2} \right. \right)$ & $(1\left|\frac{a}{2} \right. 0-\frac{c}{2} )$ \\

    $\Gamma _{1}   \left\{\begin{array}{c} {\eta _{1} } \\ {\eta _{2} } \end{array}\right. $ & $ \left[\begin{array}{cc} {1} & {0} \\ {0} & {1} \end{array}\right]$ & $ \left[\begin{array}{cc} {0} & {1} \\ {1} & {0} \end{array}\right]$ & $\left[\begin{array}{cc} {0} & {1} \\ {1} & {0} \end{array}\right]$ & $\left[\begin{array}{cc} {-1} & {0} \\ {0} & {-1} \end{array}\right]$ & $\left[\begin{array}{cc} {e^{ik_{y}b } } & {0} \\ {0} & {e^{-ik_{y}b } } \end{array}\right]$ & $\left[\begin{array}{cc} {-i} & {0} \\ {0} & {i} \end{array}\right]$ & $\left[\begin{array}{cc} {i} & {0} \\ {0} & {-i} \end{array}\right]$ \\

    $\Gamma _{2}  \left\{\begin{array}{c} {\zeta _{1} } \\ {\zeta _{2} } \end{array}\right.$ & $\left[\begin{array}{cc} {-1} & {0} \\ {0} & {-1} \end{array}\right]$ & $ \left[\begin{array}{cc} {0} & {1} \\ {1} & {0} \end{array}\right]$ & $ \left[\begin{array}{cc} {0} & {-1} \\ {-1} & {0} \end{array}\right]$ & $ \left[\begin{array}{cc} {-1} & {0} \\ {0} & {-1} \end{array}\right]$ & $ \left[\begin{array}{cc} {e^{ik_{y}b } } & {0} \\ {0} & {e^{-ik_{y}b } } \end{array}\right]$ & $ \left[\begin{array}{cc} {-i} & {0} \\ {0} & {i} \end{array}\right]$ & $ \left[\begin{array}{cc} {i} & {0} \\ {0} & {-i} \end{array}\right]$\\
  \end{tabular}
\end{table}

\begin{table}[h]
  \caption{Grey magnetic point groups (Column (b)) deduced from the minimization of the free-energy (Eq. \eqref{EQ1}). Column (a): Irreducible and reducible representations. Column (c): Equilibrium values of the order-parameters. Column (d): Spontaneous polarization components at zero magnetic fields.}\label{table2}
  \centering
\begin{tabular}{cccc}
\\
  (a)& (b) & (c) & (d) \\
  $\Gamma _{1} $ & $ 2/m1'  $ & $ \rho _{1} \ne 0,\rho _{2} =0 $ & $ P=0$ \\
  $\Gamma _{2}$ & $2/m1'$ & $\rho _{1} =0,\rho _{2} \ne 0$ & $ P=0$ \\
  \\
    & $m1'$ & $\left\{\begin{array}{c} {\rho _{1} \ne 0,\rho _{2} \ne 0} \\ {\theta _{1} -\theta _{2} =(2n+1)\frac{\pi }{2} } \end{array}\right. $ & $ P^{a} ,P^{z} $ \\
  $ \Gamma _{1} +\Gamma _{2}$ & $\bar{1}1'$ & $\left\{\begin{array}{c} {\rho _{1} \ne 0,\rho _{2} \ne 0} \\ {\theta _{1} -\theta _{2} =n\pi } \end{array}\right.$ & $P=0$ \\
    & $11'$ & $\left\{\begin{array}{c} {\rho _{1} \ne 0,\rho _{2} \ne 0}\\ {\theta _{1} -\theta _{2} \;\; arbitrary} \end{array}\right. $ & $P^{a} ,P^{b} ,P^{z} $\\
\end{tabular}
\end{table}

The equilibrium polarization is determined by minimizing the dielectric contribution to the free-energy:

\begin{equation} \label{EQ2}
F_{D}= \delta P^{a,z} \rho _{1} \rho _{2} \sin (\theta _{1} -\theta _{2} )+\frac{1}{2\chi _{0} } \left[(P^{a} )^{2} +(P^{z} )^{2} \right]
\end{equation}

where $\chi _{0}$ is the dielectric susceptibility in the paramagnetic phase. It yields $P^{a,z} =-\delta \chi _{0} \rho _{1} \rho _{2} \sin(\theta _{1} -\theta _{2})$ or equivalently $P^{a,z} =\delta (\eta _{1} \zeta _{2} -\eta _{2} \zeta _{1})$. Therefore the equilibrium form of $P^{a,z} $ in the multiferroic phase reduces to

\begin{equation} \label{EQ3}
P^{a,z} =\delta \chi _{0} \rho _{1} \rho _{2}
\end{equation}

i.e. the induced ferroelectricity has an improper character. Since the transition to the ferroic phase results from the coupling of two order-parameters it has a first-order character, $\rho _{1} $ and $\rho _{2} $ varying discontinuously across T$_{N }$. Therefore, $P^{a} $ and $P^{z}$ undergo an upward discontinuity at T$_{N}$ before increasing linearly below T$_{N}$, in agreement with the experimental curves of $P^{a,z} (T)$ measured for CuCl$_{2}$ by Seki et al.\cite{Seki2010}. The dielectric permittivity under E$^{z}$ field reported by these authors exhibits a sharp rising at T$_{N}$ before reaching a maximum at about 17K followed by a continuous decrease on cooling. Minimizing successively $F_{D} -E^{z} P^{z} $ with respect to $P^{z} $ and $E^{z}$ yields the  susceptibility component $\chi _{zz} =lim_{E_{z} \to 0} \frac{\partial P^{z} }{\partial E^{z} } $=$\chi _{0} (1-\delta \frac{\partial \rho _{1} \rho _{2} }{\partial E^{z} } )$, with  $\frac{\partial \rho _{1} \rho _{2} }{\partial E^{z} } \approx -\chi _{0} \delta (\frac{\rho _{2}^{2} }{\alpha _{1} } +\frac{\rho _{1}^{2} }{\alpha _{2} } )$ where $\rho _{1}$ and $\rho _{2}$ are the order-parameter modulus at zero field. One gets below T$_{N}$:

\begin{equation} \label{EQ4}
\chi (T)\approx \chi _{0} \left[1+2\frac{\delta ^{2} }{\Delta } \left(\frac{\beta _{1} \alpha _{2}^{2} +\beta _{2} \alpha _{1}^{2} }{\alpha _{1} \alpha _{2} }\right) -2(\gamma _{2} -\gamma _{1} )\right]
\end{equation}

where $\Delta =(\gamma _{2} -\gamma _{1} )^{2} -\beta _{1} \beta _{2} $. $\chi (T)$ increases up to $T_{Max} =\frac{a_{1} \sqrt{\beta _{2} } T_{c1} -a_{2} \sqrt{\beta _{1} } T_{c2} }{a_{1} \sqrt{\beta _{2} } -a_{2} \sqrt{\beta _{1} } }$ before decreasing continuously with T, as observed experimentally \cite{Seki2010}.

\section{Magnetoelectric interactions}

\begin{figure}[ht]
  \centering
  \includegraphics[width=\columnwidth]{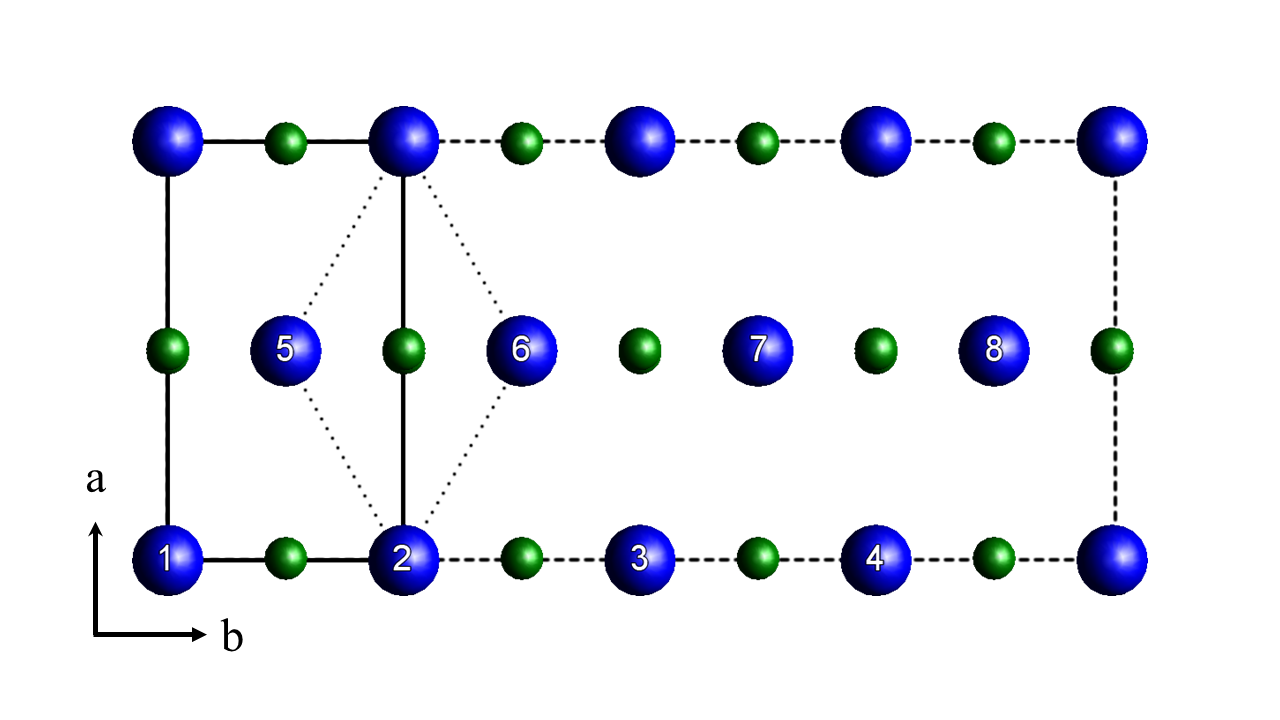}
   \caption{Conventional (solid line) and primitive (dotes line) monoclinic C-cell of CuCl$_{2}$ and CuBr$_{2}$. Eight-fold approximant (dashed line) unit-cell of the incommensurate multiferroic structure. Location of the eight copper atoms in the a-b plane. Putting the origin in atom 1, the coordinates of atoms 2-8 are respectively \textbf{b}(2), 2\textbf{b} (3), 3\textbf{b} (4), $\frac{1}{2} (\textbf{a}+\textbf{b})$ (5),$\frac{1}{2} (3\textbf{b}+\textbf{a})$ (6), $\frac{1}{2} (5\textbf{b}+\textbf{a})$ (7), and $\frac{1}{2} (7\textbf{b}+\textbf{a})$, where \textbf{b} and \textbf{c} are the lattice parameters of the paramagnetic conventional C unit-cell.
}\label{fig2}
\end{figure}

In order to determine the nature of the magnetic interactions giving rise to the electric polarization one can express the order-parameter components in function of the spin-density components $s_{i}^{u} (i=1-8,u=a,b,z)$ associated with the copper atoms of the antiferromagnetic unit-cell \cite{Banks2009}. Fig. \ref{fig2} shows the eight-fold unit-cell corresponding to an \textit{approximant} of the infinite incommensurate unit-cell in b-direction of CuCl$_{2}$ and CuBr$_{2}$. Introducing the microscopic antiferromagnetic spin-density waves \textbf{L$_{1}$= s$_{5}$-s$_{6}$+s$_{7}$-s$_{8}$, L$_{2}$= s$_{5}$-s$_{6}$-s$_{7}$+s$_{8}$}, \textbf{L$_{3}$= s$_{1}$-s$_{3}$, L$_{4}$=s$_{2}$-s$_{4}$}, one finds, by using projector techniques \cite{Toledano1987}, that the order-parameter components are spanned by the magnetic modes:

\begin{equation}
  \nonumber
   \eta _{1} (L_{1}^{a} ,L_{3}^{b} ,L_{1}^{z} ),\; \eta _{2} (L_{2}^{a} ,L_{4}^{b} ,L_{2}^{z} ), \;
   \zeta _{1} (L_{3}^{a} ,L_{1}^{b} ,L_{3}^{z} ),\; \zeta _{2} (L_{4}^{a,} ,L_{2}^{b} ,L_{4}^{z} )
\end{equation}

Assuming the magnetic moments confined in the bz plane, as reported from neutron diffraction measurements \cite{Banks2009}, the order-parameter spin waves read:

\begin{equation}
  \nonumber
  \begin{split}
  \eta _{1} =a_{1} (s_{1}^{b} -s_{3}^{b} )+a_{2} (s_{5}^{z} -s_{6}^{z} +s_{7}^{z} -s_{8}^{z} ),\eta _{2} =b_{1} (s_{2}^{b} -s_{4}^{b} )+b_{2} (s_{5}^{z} -s_{6}^{z} -s_{7}^{z} +s_{8}^{z} ) \\
  \zeta _{1} =c_{1} (s_{5}^{b} -s_{6}^{b} +s_{7}^{b} -s_{8}^{b} )+c_{2} (s_{1}^{z} -s_{3}^{z} ),\zeta _{2} =d_{1} (s_{5}^{b} -s_{6}^{b} -s_{7}^{b} +s_{8}^{b} )+d_{2} (s_{2}^{z} -s_{4}^{z} )
  \end{split}
\end{equation}

 The polarization components $P^{a,z} \approx \eta _{1} \zeta _{2} -\eta _{2} \zeta _{1} $ is formed by two types of  bilinear invariants $s_{i}^{u} s_{j}^{v} :$ 1) Antisymmetric invariants $\sum _{i\ne j}\left(s_{i}^{b} s_{j}^{z} -s_{i}^{z} s_{j}^{b} \right) $ with $(i,j)=(1-4)$ or $(5-8)$ representing the Dzialoshinskii-Moriya (DM) "cycloidal" interactions \cite{Dzyaloshinskii1965} between the spin-densities associated with atoms (1-4) or (5-8); 2) Invariants $\sum _{i,j} \left(s_{i}^{b} s_{j}^{b} \pm s_{i}^{z} s_{j}^{z} \right) $ representing anisotropic exchange interactions \cite{Whangbo2003} between the magnetic moments of atoms $i=$1 to 4 and $j=$5 to 8.

It has to be noted that taking into account the equilibrium conditions  ($\eta _{2} =0,\zeta _{1} =0$)  fulfilled by the spin-density components in the cycloidal spin-wave phase   yield:

\begin{equation}
  \nonumber
  s_{2}^{b} =s_{4}^{b} ,s_{5}^{b} -s_{6}^{b} =s_{8}^{b} -s_{7}^{b} ,s_{1}^{a,z} =s_{3}^{a,z} ,s_{5}^{a,z} -s_{6}^{z} =s_{7}^{a,z} -s_{8}^{a,z}
\end{equation}

 These conditions cancel the antisymmetric DM interactions which are replaced by anisotropic exchange interactions $\sum _{i\ne j}s_{i}^{b}  s_{j}^{z} $. It suggests that the DM interactions are inherent to the incommensurate character of the spin spiral, but are not required for the stabilization of the commensurate cycloid in the eight-fold unit-cell approximant assumed in our description.

\section{Magnetoelectric effects in $\mathbf{CuCl_{2}}$.}

Two types of magnetoelectric effects have been reported in CuCl$_{2}$ by Seki et al. \cite{Seki2010}.

1. Under increasing H$^{b}$ field the P$^{a}$ and P$^{z}$ polarization components decrease and vanish above a threshold field of about 4T at 5K. This is interpreted by Seki et al. \cite{Seki2010} as a tilting of the spin-spiral from the bc-plane to the a-c plane. Under applied H$^{b}$ field, time-reversal is broken and the magnetic point-group symmetry $m1'$ of the spin spiral phase reduces to $m$. The dependence of P$^{a}$ and P$^{z }$ on H$^{b}$ is obtained from the magnetodielectric and magnetic contributions to the free-energy which read:

\begin{equation} \label{EQ5}
F_{MP} = \frac{1}{2} \mu _{1} (M^{b} )^{2} (P^{a,z} )^{2} + \delta P^{a,z} \rho _{1} \rho _{2} \sin (\theta _{1} -\theta _{2} )
\end{equation}
\begin{equation} \label{EQ6}
F_{MH} = \frac{1}{2} \mu _{0} (M^{b} )^{2} - M^{b} H^{b}
\end{equation}

\begin{figure}[h]
  \centering
  \includegraphics[width=\columnwidth]{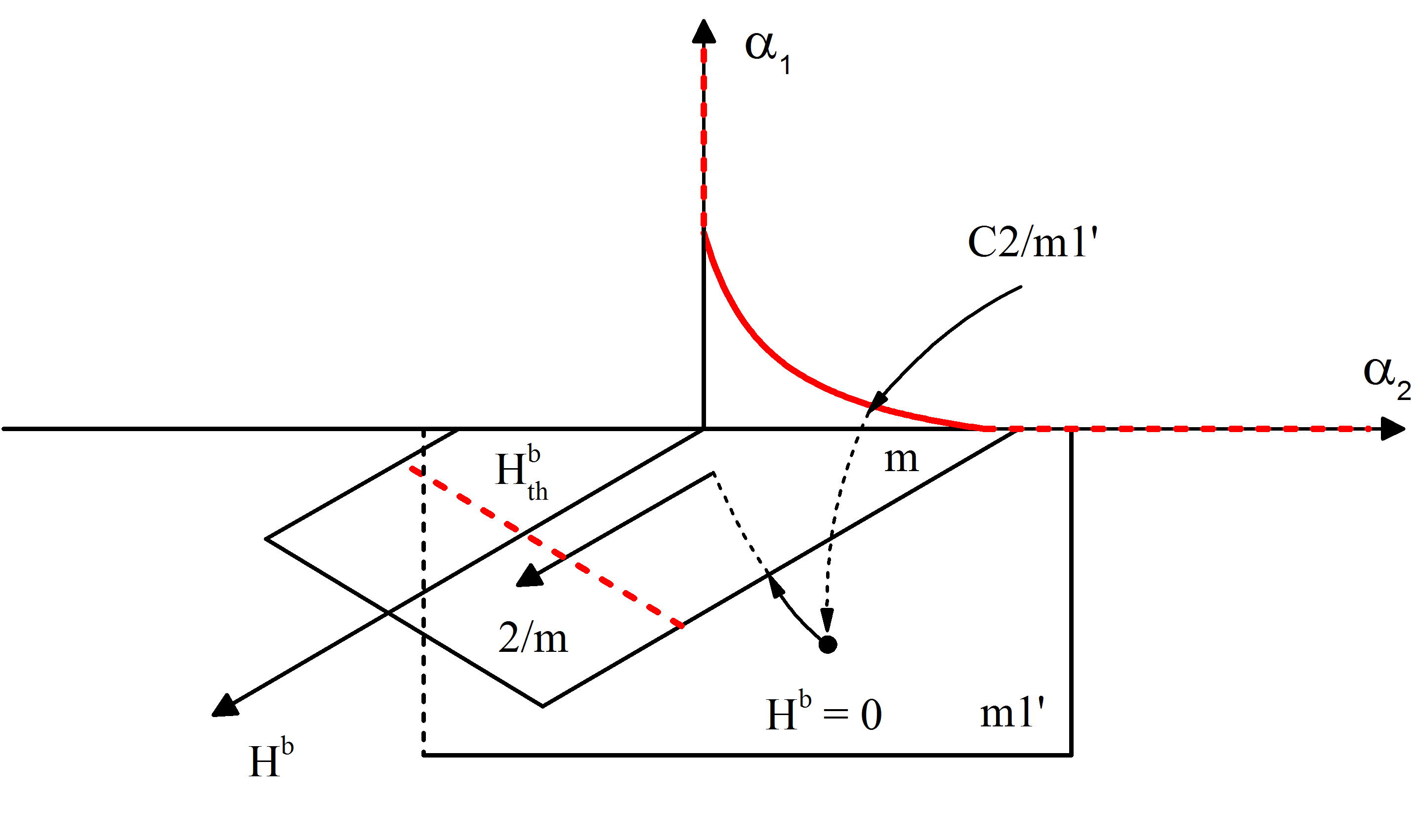}
  \caption{Theoretical phase diagram of CuCl$_{2}$ under H$^{b}$ field. The arrows show the thermodynamic path followed by the multiferroic phase with increasing fields: At low H$^{b }$ fields the m1' symmetry of the phase is lowered to m. Above a threshold field H$_{th}^{b}$ the phase loses its stability and a transition occurs towards a phase having the higher symmetry 2/m. Hatched and full red curves are, respectively, second-order and first-order transition lines.}\label{fig3}
\end{figure}

Minimizing $F_{MP}$ with respect to $P^{a,z}$, and $F_{MH}$ with respect to $M^{b}$, and putting $\theta _{1} -\theta _{2} =-\frac{\pi }{2} $ yields:

\begin{equation} \label{EQ7}
P^{a,z} (H^{b} )= \delta(\mu _{0})^2 \frac{ \rho _{1} \rho _{2} }{\mu _{1} (H^{b})^2}
\end{equation}

which shows that $P^{a}$ and $P^{z}$ decrease with increasing $H^{b}$ field.  The vanishing of the polarization components above a threshold field $H^{b} $$\approx 4$ T in CuCl$_{2}$ at 5K is obtained when $\rho _{1} \rho _{2} =0$, i. e. when $\rho _{1} $=0 or $\rho _{2} =0$. It means that above 4 T a \textit{decoupling} of the two order-parameters $\rho _{1} $ and $\rho _{2}$ occurs, only one order-parameter remaining active  Therefore the monoclinic symmetry $m$ at low field increases to $2/m$ above 4 T, consistent with the interpretation by Seki et al.$^{6}$ suggesting that the b-c spin-spiral phase becomes unstable with a tilting towards a spin-spiral located in the a-c plane. The corresponding\textit{ increase} of symmetry from $m$ to $2/m$ with increasing field realizes a field-induced ferroelectric-to-paraelectric phase transition which corresponds to the sharp peak observed for the dielectric permittivity component $\varepsilon _{zz} $ at 4 T. Fig. \ref{fig3} shows the thermodynamic path followed from the ferroelectric phase to one of the centro-symmetric phases of symmetry $2/m$ under H$^{b}$ field.

2) On rotating the magnetic field H within the a-z plane Seki et al. \cite{Seki2010} report another type of magnetoelectric effect consisting of a periodic dependence of the magnetic susceptibility and polarization components in function of the angle $\theta _{H}$ between the magnetic field and the a axis. Application of a magnetic field within the a-z plane reduces the $m1'$ symmetry at zero-field to $m'$, inducing magnetization components $M^{a}$ and $M^{z}$. The magnetic free-energy involving the couplings between the order-parameter and the magnetization components reads:

\begin{equation} \label{EQFMH}
 F_{MH} = A(\rho_i) \frac{(M^a)^2}{2} + B(\rho_i) \frac{(M^z)^2}{2} + C(\rho_i) M^a M^z - H^a M^a - H^z M^z
\end{equation}

with $A(\rho_i)=\mu^0_{aa}+\nu_{1a} \rho_1^2+\nu_{2a} \rho_2^2$, where $\mu^0_{aa}$ is the magnetic permeability and $(\nu_{1a},\nu_{2a})$ are constant coupling coefficients between the order parameter invariants $(\rho_1^2,\rho_2^2)$ and $(M^a)^2$. Analogous expressions hold for $B(\rho_i)$ and $C(\rho_i)$. Minimizing $F_{MH}$ with respect to $M^{a}$ and $M^z$ gives $M^{a} = (C H^z - B H^a)/\Delta$ and $M^{z} = (C H^a - A H^z)/\Delta$ where $\Delta= C^2 - A B$. Therefore, one has:

\begin{equation} \label{EQM}
M^{a}+M^z = \frac{(C-A)H^z+(C-B) H^a}{\Delta}= M(\cos \theta_H + \sin \theta_H)
\end{equation}

Taking into account the magnetic field dependence of 1) the order parameters, which vary as $\rho_1^2(H^{a,z}) \approx -\frac{\alpha_1}{\beta_1}-\nu_{1a}(H^a)^2-\nu_{1z}(H^z)^2-\nu_{1az} H^a H^z$ and $\rho_2^2(H^{a,z}) \approx -\frac{\alpha_2}{\beta_2}-\nu_{2a}(H^a)^2-\nu_{2z}(H^z)^2-\nu_{2az} H^a H^z$, and 2) the lowest degree approximations of $\frac{C-A}{\Delta} \approx \nu_{03}+\nu_{3a} (H^a)^2+\nu_{3z} (H^z)^2+\nu_{3az} H^a H^z$ and $\frac{C-B}{\Delta} \approx \nu_{04}+\nu_{4a} (H^a)^2+\nu_{4z} (H^z)^2+\nu_{4az} H^a H^z$, and putting $H^a=H \cos \theta_H , H^z = H \sin \theta_H$ one gets $M^a+M^z \approx (\cos \theta_H + \sin \theta_H) \left[ \lambda_1 H + H^3 (\lambda_2 + \lambda_3 \cos 2\theta_H + \lambda_4 \sin 2\theta_H) \right] $, where the $\lambda_i (i=1-4)$ are field-independent coefficients. Therefore, the total magnetic susceptibility defined by Seki et al. \cite{Seki2010} as $\chi=\frac{M}{H}=\frac{M^a+M^z}{H(\cos \theta_H + \sin \theta_h)}$ can be approximated under the form:

\begin{equation} \label{EQChi}
\chi(\theta_H) = \lambda_1 + H^2 \left( \lambda_2 + \lambda_3 \cos 2\theta_H + \lambda_4 \sin 2\theta_H  \right)
\end{equation}

\begin{figure}[h]
  \centering
  \includegraphics[width=\columnwidth]{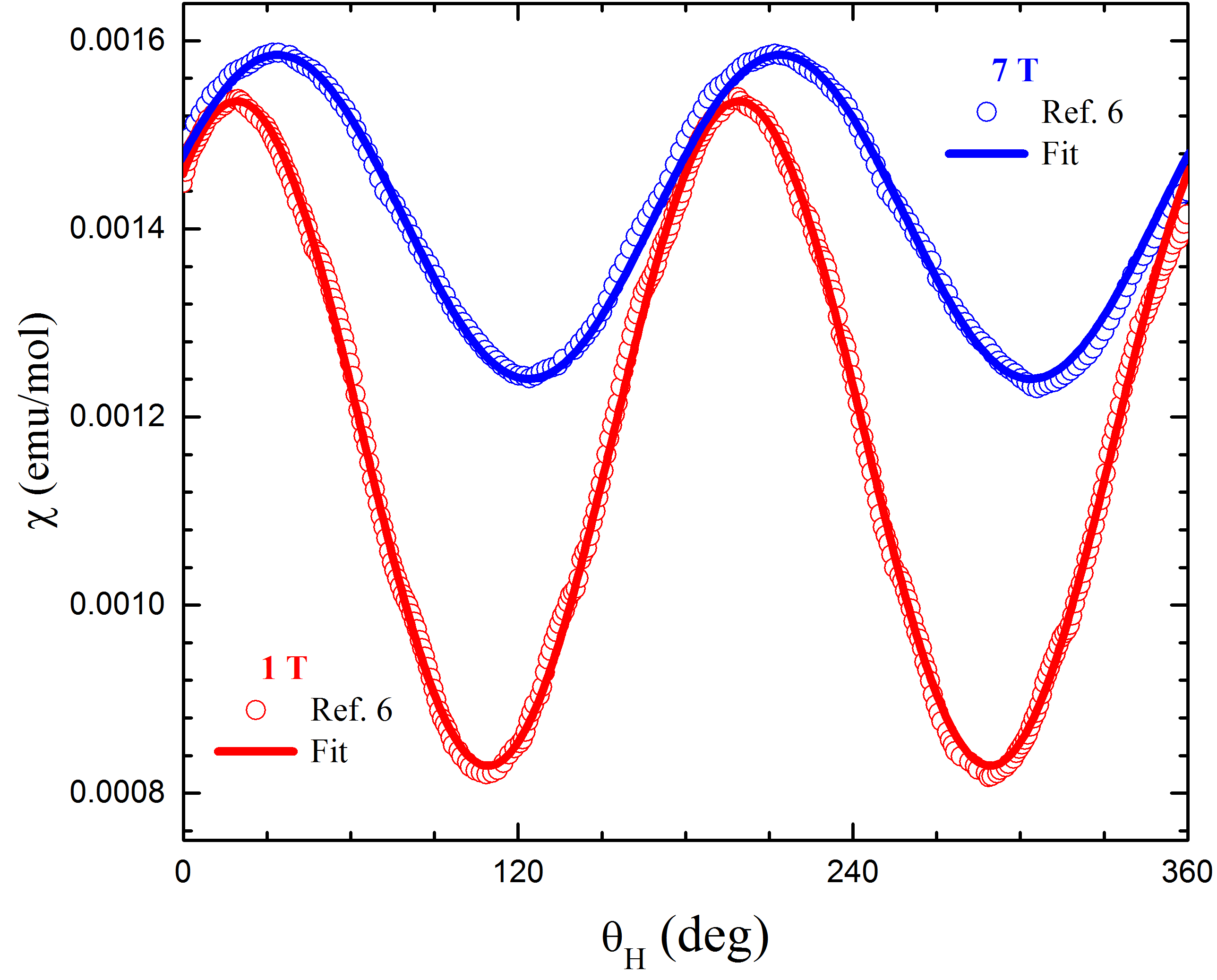}
  \caption{$\theta _{H}$-dependence of the magnetic susceptibility $\chi (\theta _{H} )$ under H$^{a,c}$ fields of CuCl$_{2 }$ following Eq.
  \eqref{EQChi}, at low (red curve) and higher field (blue curve). Experimental data was extracted from Ref. \cite{Seki2010} and fitted with the expression $\chi(\theta_H)=a + b \cos(2\theta_H) + c \cos(2\theta_H)$. The fitting parameters (in $10^{-3}emu/mol$) are for H=1T: a=1.1824(5), b=0.2773(8) and c=0.2190(8), and for H=7T: a=1.4127(6), b=0.0655(7) and c=0.1593(8)
  }
  \label{fig4}
\end{figure}

Fig. \ref{fig4} shows the periodic $\theta _{H} $-dependence of  $\chi $ for two different values of the field $H_{1} <H_{2} $, consistent with the experimental curves measured by Seki et al. \cite{Seki2010} at 1 T and 7 T.

At constant magnitude of the applied field the dependence of the polarization components on the orientation of the field can be deduced from the magnetodielectric contribution to the free-energy. For the P$^{a}$ component this contribution reads:

\begin{equation}\nonumber
F_{MP} = \frac{(P^{a})^2}{2} \left[ \frac{1}{\chi^0_{aa}} + \delta _{aa} (M^a)^2 + \delta _{az} (M^z)^2 + \delta _{aaz} M^a M^z \right] + \delta_a P^a \rho_1 \rho_2 \sin(\theta_1-\theta_2)
\end{equation}

Minimizing with respect to $P^a$ with $\theta_1-\theta_2=-\pi/2$, and taking into account the lowest-degree approximation of $\rho_1 \rho_2$ in function of the field, which is: $\rho_1\rho_2 \approx \left[ (\frac{\alpha_1\alpha_2}{\beta_1\beta_2})^2 H (\lambda_a \cos \theta_H + \lambda_z \sin \theta_H) \right]$ yields the $\theta_H$ dependence of $P^a$:

\begin{equation} \label{EQPa}
P^a (\theta_H) = \frac{\delta_a (\frac{\alpha_1 \alpha_2}{\beta_1 \beta_2})^{1/2} H (\lambda_a \cos \theta_H + \lambda_z \sin \theta_H)}
{(\chi^0_{aa})^{-1} + H^2 \left[\delta _{aa} \cos^2 \theta_H + \delta _{az} \sin^2 \theta_H + \delta _{aaz} \sin \theta_H \cos \theta_H \right]}
\end{equation}

and an analogous expression for $P^z (\theta_H)$:

\begin{equation} \label{EQPb}
P^z (\theta_H) = \frac{\delta_z (\frac{\alpha_1 \alpha_2}{\beta_1 \beta_2})^{1/2} H (\lambda_a \cos \theta_H + \lambda_z \sin \theta_H)}
{(\chi^0_{zz})^{-1} + H^2 \left[\delta _{zz} \cos^2 \theta_H + \delta _{za} \sin^2 \theta_H + \delta _{zza} \sin \theta_H \cos \theta_H \right]}
\end{equation}

\begin{figure}[h]
  \centering
  \includegraphics[width=\columnwidth]{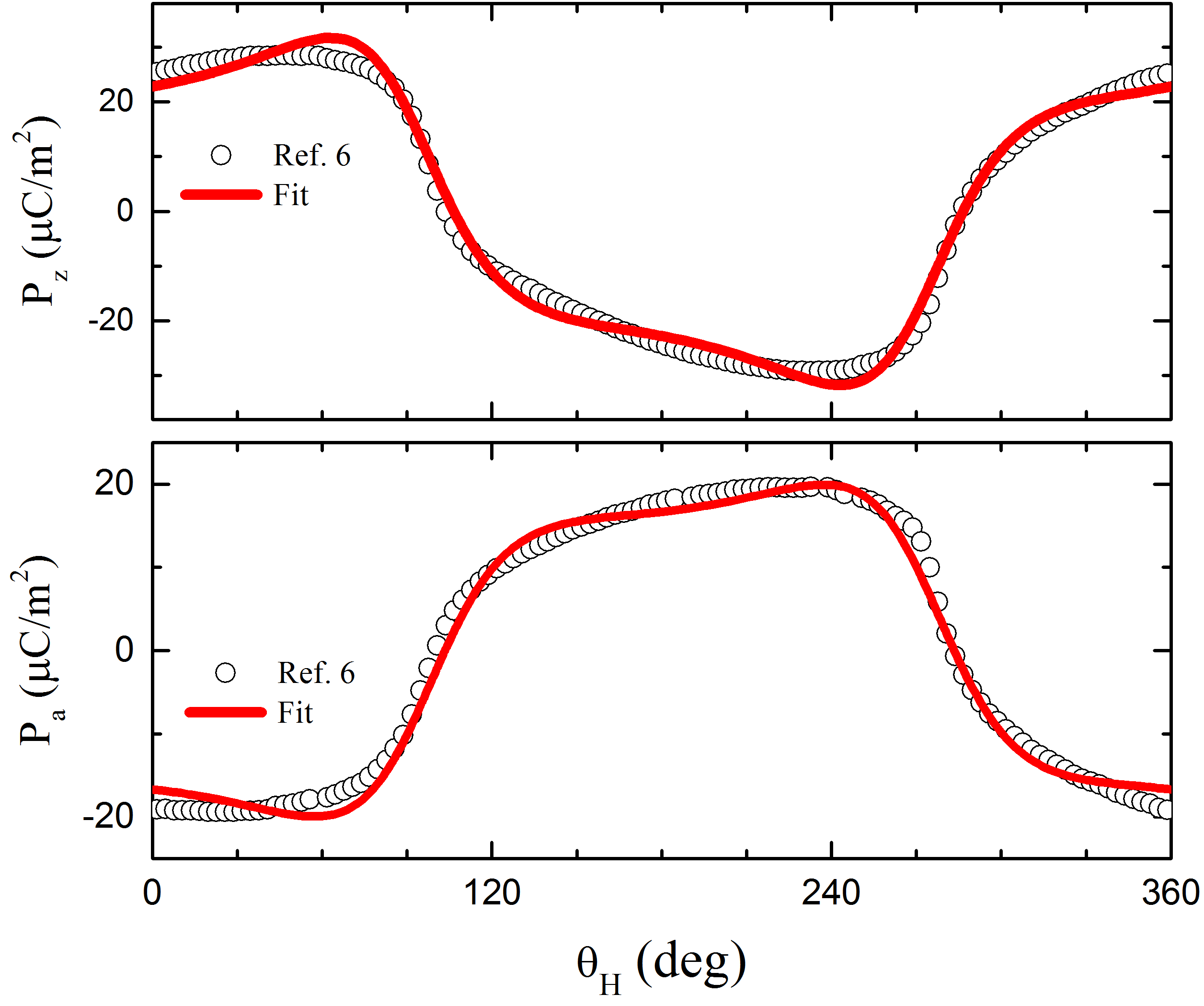}
  \caption{$\theta _{H}$-dependence of the P$^{a}$ and P$^{z}$ polarization components of CuCl$_{2}$ according to Eqs. \eqref{EQPa} and \eqref{EQPb}, respectively. Experimental data was extracted from Ref. \cite{Seki2010} and fitted with the expression $P^{a,z}(\theta_H)= a(\cos \theta_H + b \sin \theta_H)/(1 + c \cos^2\theta_H + d \cos \theta_H \sin \theta_H)$. The fitting parameters (a in $\mu C/m^2$) are for $P^z$: a=62.6(4), b=0.298(2), c=1.75(3) and d=0.14(2), and for $P^a$: a=-42.3(5), b=0.238(2), c=1.53(4) and d=0.27(3). }
  \label{fig5}
\end{figure}

Fig. \ref{fig5} shows the periodic dependence on $\theta _{H}$ of P$^{a}$ and P$^{z}$ which coincide with the experimental curves reported by Seki et al. \cite{Seki2010}. Reversing the field by $180^\circ$ changes $\theta_H$ into $\theta_H + \pi$, reversing the sign of $P^a$ and $P^z$ as observed experimentally \cite{Seki2010}.

\section{ Magnetoelectricity in $\mathbf{CuBr_{2}}$}

Magnetoelectric properties have been also evidenced in CuBr$_{2}$\cite{Zhao2012}  below  T$_{N}$ = 73.5 K. As for the isostructural compound CuCl$_{2}$, neutron powder diffraction patterns  by Zhao et al. \cite{Zhao2012} confirm the existence of an incommensurate helical spin spiral propagating along the b axis, with an approximate quadrupling of the nuclear unit-cell along b and doubling along c. Therefore, the symmetry analysis and phase diagram proposed for CuCl$_{2}$ is also valid for CuBr$_{2}$, i.e. the observed emergence of an electric polarization below T$_{N}$ results from the coupling of the same bi-dimensional order-parameters associated with the reducible representation $\Gamma _{1} +\Gamma _{2}$ (Table \ref{table1}).  However, since the dielectric measurements have been performed on polycrystalline samples of CuBr$_{2}$ the intrinsic anisotropy of the dielectric properties is unclear. Thus, if at zero field the symmetry of the incommensurate helical spin spiral corresponds to the grey point-group $m1'$, one cannot deduce from the dielectric measurements the symmetry of the phase in presence of magnetic field which is lowered to $m$ for a H$^{b}$ field, to $m'$ for H$^{a}$ or H$^{z}$ fields and to $1$ for a field in general direction. Although a detailed theoretical description of the magnetoelectric effects in CuBr$_{2}$ cannot be presently achieved, the following analysis can be made of the dielectric and magnetoelectric observations reported by  Zhao et al. \cite{Zhao2012}

Two correlated magnetoelectric effects are described by these authors: An increase of the total polarization with increasing magnetic field, the saturated value of P(H) increasing from 8 $\mu C/m^{2}$ at 0 T to 22.5 $\mu C/m^{2}$ at 9 T, and a decrease of the maximum dielectric permittivity $\varepsilon _{\max }(T)$ with increasing H. The first effect can be derived from the magneto-dielectric free-energy $F_{PM} =$$\frac{P^{2} }{2\chi _{0} } +\delta PM^{2} \rho _{1} \rho _{2}$ and magnetic free-energy $F_{M} =\mu _{0} \frac{M^{2} }{2} +M^{2} (\delta _{1} \rho _{1}^{2} +\delta _{2} \rho _{2}^{2} )-MH$ which yield:

\begin{equation} \label{EQ13}
P(H)=\frac{\chi_0 \delta \rho_1 \rho_2 H^2 }{(\mu_0 +\delta_1 \rho_1^2 +\delta_2 \rho_2^2)^2}
\end{equation}

Corresponding to a increase of the maximum polarization with H. Eq. \eqref{EQ13} is also consistent with the saturated regime of $P(T)$ observed on cooling from 60 K to 10 K. Note in this respect that the absence of decrease of $P(T)$ down to 10 K suggests that a single ferroelectric phase is stabilized on cooling below T$_{N}$ in CuBr$_{2}$.

A sharp rising of the dielectric permittivity $\varepsilon (T)$ at T$_{N}$, followed by a continuous decrease with decreasing temperature is observed at zero field in CuBr$_{2}$ \cite{Zhao2012} as for CuCl$_{2}$ \cite{Seki2010} which follows Eq. \eqref{EQ4}. Under applied magnetic field the dielectric susceptibility in the multiferroic phase reads:

\begin{equation} \label{EQ14}
\chi (H)=\chi (0)+D(\rho _{i} ) \left(1-\frac{H^{2} }{\left(\mu _{0} +\delta _{1} \rho _{1}^{2} +\delta _{2} \rho _{2}^{2} \right)^{2} } \right)
\end{equation}

where $D(\rho _{i} )=-\chi _{0}^{2} \delta ^{2} \left(\frac{\rho _{2}^{2} }{\alpha _{1} } +\frac{\rho _{1}^{2} }{\alpha _{2} } \right)>0$ for $\alpha _{1} <0$ and $\alpha _{2} <0$. Therefore with increasing field the dielectric susceptibility decreases, as reported in Ref. \cite{Zhao2012}.

\section{Summary, discussion and conclusion.}

In summary, our proposed theoretical analysis shows that the multiferroic phase transitions occurring in CuCl$_{2}$ and CuBr$_{2}$ are induced by the coupling of two distinct antiferromagnetic spin-wave order parameters which lead, across a first-order transition, to an incommensurate polar phase of monoclinic symmetry $m1'$ displaying a typical improper ferroelectric behaviour. The set of spin-density components spanning the order-parameters have been worked out. They indicate that the emergence of an electric polarization results from combined Dzialoshinskii-Moriya (DM) antisymmetric interactions and symmetric anisotropic exchange interactions. Interestingly, the antisymmetric interactions are cancelled when assuming a commensurate approximant of the magnetic structure instead of the actual incommensurate structure. It indicates that the DM interactions which are required for stabilizing the spiral structure and suppressing the inversion symmetry, contribute only partly to the electric polarization. This conclusion differs from the interpretation by Seki et al. \cite{Seki2010} of the dielectric properties of CuCl$_{2}$ in terms of an exclusive inverse DM interaction \cite{Katsura2005}.

The different magnetoelectric effects observed in CuCl$_{2}$ and CuBr$_{2}$ under applied magnetic fields have been described theoretically by considering the couplings existing between the order-parameters, the polarization and the magnetization. In particular, the disappearance of the polarization above a threshold magnetic H$^{b}$ field has been shown to result from a decoupling of the order-parameters which are coupled at zero fields. It suggests that the coupling of the spin-waves inducing the antiferromagnetic spin-spiral is weak and sensitive to the applied field.

Although directional magnetoelectric effects on single crystal of CuBr$_{2}$ remain to be investigated in more details, the effects reported by Zhao et al. \cite{Zhao2012} from powder sample measurements show a remarkable coincidence with those observed in CuCl$_{2}$, despite a difference of about 50 K for their transition temperatures: The same spin-wave order-parameter symmetries are activated in CuCl$_{2}$ and CuBr$_{2}$, involving the same interactions between the spin-densities and similar couplings between the induced polarization and magnetization components.  This coincidence is due to the similarity of their structures, consisting of undistorted triangular lattices involving halide ions. However the presence of copper ions should also be important in the determination of the specific multiferroic properties in CuCl$_{2}$ and CuBr$_{2}$. In this respect, it can be noted that cupric oxide \cite{Kimura2008,Toledano2011a} is one among the few multiferroic oxides in which a direct transition to a spiral multiferroic phase is observed, as it is the case for the two copper halides.

\begin{acknowledgments}
The authors thank CNPq (Brazilian Federal Funding Agency) for financial support.
\end{acknowledgments}


\end{document}